\newcommand{\version}{27/05/04}
\documentclass[prl,aps, preprint,
showpacs]{revtex4}
\usepackage{amsgen,amstext,amsbsy,amsopn,amsthm, amssymb}
\usepackage{amsmath}

\begin{document}

\title{String-localized quantum fields from Wigner representations}

\author{Jens Mund}
\email{mund@fma.if.usp.br}
\affiliation{Instituto de F\'{\i}sica, Universidade de 
S\~{a}o Paulo, 
CP 66\,318, 05315 - 970 S\~{a}o Paulo, SP, Brazil}
\author{Bert Schroer}
\email{schroer@physik.fu-berlin.de}
\affiliation{CBPF, Rua Dr.\ Xavier Sigaud 150, 22290-180 Rio de 
Janeiro, Brazil, and \\ Institut f\"ur Theoretische Physik, FU-Berlin, 
Arnimallee 14, D-14195 Berlin, Germany}
\author{Jakob Yngvason}
\email{yngvason@thor.thp.univie.ac.at}
\affiliation{Institut f\"ur Theoretische Physik,
Universit\"at
Wien, Boltzmanngasse 5, A-1090 Vienna, Austria, and\\ 
Erwin Schr\"odinger Institute for Mathematical Physics,
 Boltzmanngasse 9, A-1090 Vienna, Austria}

\date{\version}

\begin{abstract}
In contrast to the usual representations of of the Poincar\'e group of
finite spin or helicity the Wigner representations of mass zero and
infinite spin are known to be incompatible with pointlike localized
quantum fields.  We present here a construction of quantum fields
associated with these representations that are localized in
semi-infinite, space-like strings. The construction is based on
concepts outside the realm of Lagrangian quantization with the
potential for further applications.
\end{abstract}

\maketitle It is well-known that free fields for particles of finite
spin (or helicity in case of $m=0$) can be constructed in two ways,
either by (canonical or functional integral) Lagrangian quantization,
or within the setting of Wigner's particle classification \cite{Wig1}
based on positive energy representations of the universal covering of
the Poincar\'{e} group \cite{Weinberg}.  There is, however, a family
of representations where the standard field-theoretical procedures
fail.  These representations correspond to particles of zero mass and
infinite spin and can be regarded as limiting cases of representations
of mass $m>0$ and spin $s<\infty$ as $m\to 0$ and $s\to\infty$ with
the Pauli-Lubanski parameter $m^2s(s+1)=\kappa^2$ fixed and nonzero. 
In the Wigner classifications they are associated with faithful
representations of the noncompact stabilizer group (``little group'')
of a light-like vector.  In this case no Lagrangian description is
known; in fact there exists a No-Go theorem \cite{Yngvason} stating
that these representations are incompatible with pointlike localized
fields fulfilling the general principles of quantum field theory
\cite{Streater-Wightman}.  Special examples that indicate the
difficulties to make these representations compatible with the
structure of local fields can also be found in \cite{Mack-Iverson,
Abbot}.

In this Letter we report on the construction of {\it string-localized}
fields for these representations; the string turns out to be a
semi-infinite space-like line characterized by an initial point $x$ in
Minkowski space and a space-like direction $e$ from the unit
space-like hyperboloid (a point in a de Sitter space).  In this paper
\lq\lq localization" is always understood in terms of the vanishing or
nonvanishing of commutators of field operators, and
string-localization means that the commutator of two field operator
vanishes if the corresponding strings are space-like separated but in
general not if this holds only for the end points.  The existence of
string-localized objects as the best possible (with the tightest
localization) for these representations is suggested by recent general
results on localization in space-like cones that apply to all positive
energy representations of the Poincar\'{e} group \cite{BGL}.  Our
string-localized fields transform in a simple way under the
Poincar\'{e} group and their internal degrees of freedom consist in
the infinite helicity tower of a faithful representation of the
Euclidean stabilizer group $E(d-2)$ in space-time dimension $d\geq 4$. 
For $d=3$ the representation is
one-dimensional but leads also to string-localized fields.  For
concreteness sake we consider here the case $d=4$ and integer 
helicities.  Our findings solve
an old problem that has attracted the attention of physicists of
several generations \cite{Wig2, Yngvason, Mack-Iverson, Abbot}, namely
to incorporate these representations into quantum field theory in a
way compatible with causality.

New concepts, outside the realm of Lagrangean quantization, have been
essential for our construction. We regard our work as an argument in
favour of the strength and relevance to QFT of these concepts, which
have the potential for further applications as pointed out below. 

An interesting feature of our construction is a subtle interplay
between the pointlike localization of the end point of the string in
$d$-dimensional Minkowski space and the directional localization in a
$(d-1)$-dimensional de Sitter space in the sense of
\cite{Bros-Moschella}.  We note that in his search for a classical
local equation for the zero mass infinite spin representations Wigner
\cite{Wig2} proposed a description in which the Poincar\'e group also
acts on a space-like vector besides the points in Minkowski space. 
The wave equations of \cite{Wig2}, however, are inconsistent with
string-localization in the sense considered here.

The infinite spin Wigner representations are not the only irreducible
representations leading to string localization; massive
representations in $d=1+2$ with spin not equal to an integer or
half-integer (anyons) can only be string localized.  In that case the
string localization results from the richer covering structure of the
$d=1+2$ Poincar\'e group which also leads to braid group statistics
which requires the presence of vacuum polarization even in the absence
of a genuine interaction (absence of real particle creation)
\cite{Mund2}.  The anyonic string is a special case of the string-like
localized objects envisaged in \cite{BuFr}.

In this context it is worth pointing out that there is a significant
difference between string localization in our sense and localization
in string field theory.  The lightfront quantization of the free
bosonic Nambu-Goto string leads according to the analysis in
\cite{dimock} to pointlike localization in the sense that the
commutator vanishes for space-like separation of the centers of mass
of two string configurations, irrespective of an overlap of their
internal coordinates.  For interacting string field theory there are
no rigorous results of this kind, but perturbative calculations
\cite{L, LL} seem to indicate that if such a theory is meaningful at
all (which is by no means clear) the string fields can be expected to
be totally delocalized.  On the available evidence it seems in any
case fair to say that the strings of string field theory are not
string-localized in the sense of the present paper.

Our construction of string-localized fields is based on Tomita-Takeski
modular theory (see \cite{Borchers} for a survey of its applications
to quantum field theory) in the context of modular localization for
Poincar\'{e} covariant positive energy representations \cite{BGL,
Mund1, S99, F-S, BuL}.  A full treatment in the modular setting will
be given in \cite{M-S-Y}.  Here we only describe the main result and give an
argument which (in the present condensed version) is less systematic
and rigorous but has the advantage of being more accessible to readers
with a standard field-theoretic background. For convenience of the 
reader we include some basic definitions and facts about modular 
localization in an appendix.

We start our construction by recalling the definition of the
irreducible zero mass, infinite spin representations of the proper,
orthochronous Poincar\'e group ${\mathcal P}^{\uparrow}_+$.  They are
defined by inducing unitary representations of the stabilizer group of
a fixed light-like vector to the whole of ${\mathcal P}^{\uparrow}_+$. 
The stabilizer group is in our case isomorphic to the two-dimensional
Euclidean group $E(2)$, consisting of rotations $R_{\vartheta}$ by an
angle $\vartheta\in\mathbb{R}{\,\rm mod \,}2\pi$ and translations by
$c\in\mathbb{R}^2$.  Let ${\mathcal H}_{\kappa}$ be the Hilbert space
of functions of $k\in\mathbb{R}^2$, square integrable with respect to
the measure $d\nu_{\kappa}=\delta(\left| k\right| ^{2}-\kappa
^{2})d^2k$.  (Hence only the restrictions of the functions to a circle
of radius $\kappa$ matter.)  The Pauli-Lubanski parameter $\kappa^2$
labels nonequivalent representations of $E(2)$; the representation on
${\mathcal H}_{\kappa}$ is given by the formula
\begin{equation}\label{formula}\left(
D_{\kappa}(c,R_{\vartheta})\varphi\right) (k)=e^{ic\cdot
k}\varphi(R_{\vartheta} ^{-1}k).\end{equation}

Let $\psi(p)$ be an ${\mathcal H}_{\kappa}$-valued wave function of
$p\in\mathbb{R}^4$, square integrable with respect to the Lorentz
invariant measure $d\mu(p)=\theta(p^0)\delta(p\cdot p)d^4 p$ on the
mantle $\partial V^+$ of the forward light cone $V^+$.  The unitary
Wigner transformation law for such a wave function reads
\begin{equation}
\left(U(a,\Lambda)\psi\right)(p)=e^{ip\cdot a}D_{\kappa}(R(\Lambda,p))
\psi(\Lambda
^{-1}p)\label{wave}
\end{equation}
where \begin{equation}\label{wignerrot}
R(\Lambda,p)=B_{p}^{-1}\Lambda B_{\Lambda^{-1}p}\in E(2)\end{equation}
denotes the Wigner ``rotation'' (actually a boost combined with a 
rotation) with $B_{p}$  an appropriately chosen
Lorentz transformation that transforms the standard vector $\bar 
p=(1,0,0,1)$ to  a (nonzero) $p\in \partial V^+$.

Our string-localized field operators are defined on the Fock-space
over the irreducible representation space with the creation and
annihilation operators $a^*(p)(k)$, $a(p)(k)$ for the basis kets
$|p,k\rangle$ of the one-particle space, $p\in\partial V^+$,
$k\in\mathbb{R}^2$, $|k|=\kappa$.  In fact, we define a whole family
of fields, depending on a complex parameter $\alpha$ that labels
representations of the 3-dimensional de Sitter group as will be
explained in the sequel.  The field operators have the form
\begin{eqnarray}
\Phi^\alpha(x,e)=\int_{\partial V^{+}}d\mu(p)\left\{  
e^{ipx}u^{\alpha}(p,e)\circ
a^{\ast}(p)\right.\nonumber\\ \left.
+e^{-ipx}\overline{u^{\bar{\alpha}}(p,e)}\circ a(p)\right\}
\end{eqnarray}
with ${\mathcal H}_{\kappa}$-valued prefactors $u^\alpha(p,e)$ that
are determined by the intertwining property \eqref{intertwine} below
and certain analyticity requirements for their dependence on $e$.  The
circle ``$\circ$'' between the prefactors $u^\alpha(p,e)$ and the
creation and annihilation operators (the dependence on $k$ is
suppressed by the notation) stands for integration over $k\in
\mathbb{R}^2$ with respect to the measure $d\nu_{\kappa}(k)$, and the
bar denotes complex conjugation.  The fields are singular in $x$ and
the space-like direction $e$, i.e., operator valued distributions, and
they have the following properties:
\begin{itemize}
\item If $x+\mathbb{R}^{+}e$ and $x^{\prime}+\mathbb{R}^{+}e^{\prime}$
are space-like separated \footnote{The distributional 
character of the fields requires in fact strict separation in the sense that 
that some open neighborhoods of the strings are 
space-like separated.} then
\begin{equation}\label{loc}
\left[  \Phi^\alpha(x,e),\Phi^{\alpha'}(x^{\prime},e^{\prime})\right]  =0
\end{equation}
while the commutator is nonzero as a distribution in
$e,e^\prime$ if only the the endpoints of the strings,
$x$ and $x'$, are
space-like separated.

\item  The transformation law of the field is consistent with this localization:
\begin{equation}\label{cov}
U(a,\Lambda)\Phi^\alpha(x,e)U(a,\Lambda)^{-1}=\Phi^\alpha(\Lambda x+a,\Lambda e).
\end{equation}

\item After smearing with tests functions in $x$ and $e$, where it is
sufficient to let $x$ and $e$ vary in an arbitrary small region, the
field operators generate a dense set in Fock space when applied to the
vacuum vector $|0\rangle$.  (Reeh-Schlieder property
\cite{Streater-Wightman}.)
\end{itemize}

The second statement \eqref{cov} is a result (as in the standard
finite spin case) of the intertwining properties of $u^{\alpha}$,
namely $u^{\alpha}$ and $\overline{u^{\bar{\alpha}}}$ absorb the
Wigner rotation of the creation/annihilation operators (which is
contragradient to that of the wave function \eqref{wave}) and trade it
for a transformation of $e$ according to \begin{equation}
\label{intertwine} D_{\kappa}(R(\Lambda,p))u^{\alpha}
(\Lambda^{-1}p,e)=u^{\alpha}(p,\Lambda e).\end{equation} The
localization \eqref{loc}, on the other hand, results from \eqref{cov},
TCP covariance, and analyticity properties of the two point function
in $x-x'$ and in $e$, $e'$.  The third property is proved in a similar
way as the Reeh-Schlieder theorem for point-localized fields
\cite{Streater-Wightman}, using also analyticity in $e$.  The field
operators for different values of the parameter $\alpha$ all generate
the same Fock space and Eq.\ \eqref{loc} implies that they are
relatively (string) localized to each other.  Hence they all belong to
the same Borchers class \cite{Borchersclass}.

The intuitive basis of this construction is the idea that one can
obtain the relevant representation by a suitable projection from a
tensor product representation, where one factor is a scalar massless
Wigner representation of the Poincar\'e group in $d=4$ dimensional
Minkowski space and the other a representation of the Lorentz group
associated with a $d-1=3$ dimensional de Sitter space.  Without any
relation between the tensor factors, one would obtain a factorizing
two-point function associated with a commutator that vanishes if both
the Minkowski- and de Sitter localizations points are space-like.  The
action of the Poincar\'{e} group in the tensor product space
$\mathcal{H}=\mathcal{H}_{0}\bar{\otimes
}\mathcal{H}_{{\rm dS}}$ is $U_{{\rm tens}}(a,\Lambda)=U_{0}(a,\Lambda)\bar{\otimes}%
U_{\rm dS}(\Lambda),$ where $U_{0}(a,\Lambda)$ is the Wigner
representation of a massless, scalar particle, and $U_{{\rm
dS}}(\Lambda)$ is a representation of the homogenous Lorentz-group on
functions on $d-1$ dimensional de Sitter space as in
\cite{Bros-Moschella} of degree $\alpha$, which is unitary if
$\alpha=-\frac{d-2}{2}+i\rho,$ $\rho\in$ $\mathbb{R}$ \footnote{A
closely related use of representations of the homogeneous Lorentz
group is made in \cite{Mack-Iverson}.  The essential difference is
that string-localization, which is our main concern, is not visible in
this earlier construction.}. It turns out that for our purpose all
values of $\alpha$ are allowed (except $\alpha=0,1,2,\dots$ for which
$u^\alpha\equiv 0$ for $k\neq 0$ by Eqs.\ \eqref{ualpha} and 
\eqref{xi} below), 
but the unitary case, ${\rm
Re}\,\alpha=-1$, is perhaps the most natural choice.  For unitary
$U_{\rm dS}$ the representation $U_{{\rm tens}}(a,\Lambda)$ is a direct integral of
the continuum of infinite spin Wigner representations corresponding to
all real values of the Pauli-Lubanski parameter $\kappa$.  Projecting
out one of these uncountably many irreducible representations weakens
the independent localizations in $x$ and $e$ in such a way as to be
consistent with the mutual causal dependency of strings.  The
decomposition of the tensor product representation into its
irreducible components is carried out by first bringing it into the
Wigner form (i.e., the form of \eqref{wave}) by means of a unitary
transformation $\psi(p)\to U_{\rm dS}(B_{p})\psi(p)$ and then
decomposing it according to the spectrum of the Casimir operator of
the little group. A definite value of $\kappa$ is then picked out.
The resulting intertwiners are
\begin{equation}\label{ualpha}
u^{\alpha}(p,e)(k) =
e^{-i\pi\alpha/2}
\int d^{2}ze^{ik\cdot z}\left(  B_{p}
\xi(z)\cdot e\right)  ^{\alpha}
\end{equation}
with
\begin{equation}\label{xi}
\xi(z)    =\left({\mbox{$\frac{1}{2}$}}\left(  \left|  z\right|  ^{2}+1\right)
,z_1,-z_2,{\mbox{$\frac{1}{2}$}}\left(  \left|  z\right|
^{2}-1\right)  \right).
\end{equation}
Here $\xi\in\partial V^{+}$ is a de Sitter momentum space variable,
and $\left( \xi\cdot e\right) ^{\alpha}$ (the dot denotes here the
Minkowski inner product) is the analog of a plane wave, i.e., as a
function of $\xi$ and the exponent $\alpha$ it is the Fourier-Helgason
transform of the $\delta$-function at the point $e$ in de Sitter space
as explained in \cite{Bros-Moschella}.  The power $t^\alpha$ is
defined with a cut along $\mathbb{R}^-$ and
$(-1)^\alpha=\exp(i\pi\alpha)$.  Instead of integrating $\xi$ over
time-like or space-like cycles $\Gamma$ as the authors of
\cite{Bros-Moschella}, we chose the light-like cycle
$\Gamma_{(1,0,0,1)}=\left\{ \xi\in\partial V^{+},\left( \xi\cdot
e\right) =1\right\} $ that leads to the parametrization \eqref{xi} in 
terms of points
$z\in \mathbb{R}^2$.  The integral in \eqref{ualpha} is understood in
the sense of tempered distributions, but by partial integration one
sees that for $k\neq 0$ the result is a continuous function of 
$k\in{\mathbb R}^2$ that can be restricted to $|k|=\kappa$.

Since $B_{p}\xi(z)\in\partial V^+$ has a positive scalar product with
any vector in the forward light cone $V^+$, it follows from
\eqref{ualpha} that $u^{\alpha}(p,e)(k)$ can be defined for complex
vectors $e$, provided the imaginary part of $e$ is in $V^+$. 
Moreover, $u^{\alpha}(p,e)(k)$ is analytic in $e$ in this domain.

The nontrivial coupling between initial points and directions arises
from the presence of the $p$-dependent boost $B_{p}$ and of the 2D
plane wave factor $e^{ik\cdot z}$ which produces the variable $k$ on
which the Lorentz group acts through the Wigner ``rotation''
$D_{\kappa}(R(\Lambda ,p))$, c.f.\ \eqref{formula} and
\eqref{wignerrot}.  This action, consisting of a two-dimensional
translation $c$ and a rotation $R_{\vartheta}$ both depending on
$\Lambda$ and $p$ (i.e., $R(\Lambda,p)=(c,R_{\vartheta})),$ can be
pulled through to the $z$ in $\xi(z)$ as follows:
\begin{align}\label{intertwining}
&  D_{\kappa}(R(\Lambda,p))u^{\alpha}(\Lambda^{-1}p,e)(k)=e^{ic\cdot k}u^{\alpha
}(\Lambda^{-1}p,e)(R_{\vartheta}^{-1}k)\,\nonumber\\
&=
e^{-i\pi\alpha/2}
\int d^{2}ze^{ik\cdot z}\left(  B_{\Lambda^{-1}p}R(\Lambda,p)^{-1}\xi\left(
z\right) \cdot e\right)  ^{\alpha}\nonumber\\
&  =
e^{-i\pi\alpha/2}
\int d^{2}ze^{ik\cdot 
z}\left(  \Lambda^{-1}B_{p}\xi\left(
z\right) \cdot e\right)  ^{\alpha}=u^{\alpha}(p,\Lambda e)(k),
\end{align}
verifying \eqref{intertwine}. Here we have in the second line used the relation $\xi(R_{\vartheta}%
z+c)=R(\Lambda,p)\xi(z)$ that follows directly from the above formula
\eqref{xi} for $\xi(z).$ The passing to the third line uses the
formula \eqref{wignerrot} for Wigner rotation $R(\Lambda,p)$.  Besides
the representation of ${\mathcal P}^\uparrow_{+}$ an antiunitary TCP
transformation is defined by $|p,k\rangle\to |p,-k\rangle$, which
means that $u^\alpha(p,e)(k)\to\overline{u^\alpha(p,e)(-k)}=
{u^{\bar\alpha}(p,-e)(k)}$.  This sets the stage for the application
of the modular localization \cite {BGL} of one-particle states that
can be shown to imply the desired string commutation relation.  We
shall not discuss this approach here but pass directly
to the commutator via the two-point function%
\begin{align}\label{twopoint}
\mathcal{W}^{\alpha\alpha'}(x-x';e,e^{\prime})  &  =\left\langle 
0|\Phi^\alpha(x,e)\Phi^{\alpha'}(x^{\prime
},e^{\prime})|0\right\rangle\nonumber\\ &=
\int_{\partial V^{+}}d\mu(p)e^{-ip\cdot (x-x')}M^{\alpha\alpha'}
(p;e,e^{\prime}),\\
M^{\alpha\alpha'}(p;e,e^{\prime})  &  
=\overline{u^{\bar{\alpha}}(p,e)}\circ u^{\alpha'}%
(p,e^{\prime}),\nonumber
\end{align}
where $\circ$ again denotes integration over $k$ on the circle 
$|k|=\kappa$. In contradistinction to pointlike localized fields, 
where $M^{\alpha\alpha'}$ 
is a polynomial in $p$,  we cannot express this two-point
function in terms of known functions but we can read off its covariance
properties from 
Eq.\ \eqref{intertwine} and the TCP symmetry in the one-particle space:
\begin{align}
M^{\alpha\alpha'}(p;\Lambda e,\Lambda e^{\prime})  &  
=M^{\alpha\alpha'}(\Lambda^{-1}p;e,e^{\prime})\label{cov1}\\
M^{\alpha\alpha'}(p;-e,-e^{\prime})  &  =M^{\alpha'\alpha}( 
p;e^{\prime},e).\label{cov2}
\end{align}

Since the measure $d\mu(p)$ has support on $\partial V^+$ the
two-point function $\mathcal{W}^{\alpha\alpha'}(x-x';e,e^{\prime})$ is
an analytic function of $x-x'$ in the complex domain
$\mathbb{R}^4-iV^+$.  Moreover, by the analyticity of $u^\alpha$ in
$e$, $\mathcal{W}^{\alpha\alpha'}$ is analytic for complex $e'$ with
$e'\cdot e'=-1$ and imaginary part in $V^+$.  Likewise, it is
antianalytic for complex $e$ in the same domain.

If two strings, $x+\mathbb{R}^+ e$ and $x'+\mathbb{R}^+ e'$ are
space-like separated (cf.\ footnote [22]), 
there is a space-like wedge $W$ with causal
complement $W'$ such that $x+\mathbb{R}^+ e\in W$ and $x'+\mathbb{R}^+
e'\in W'$.  By translational invariance of the two-point function it
can be assumed that the edge of $W$ (and hence also of $W'$) contains
the origin; then $x,e\in W$ and $x',e'\in W'$.  The covariance law
\eqref{cov1} and the TCP symmetry \eqref{cov2} imply the following
``exchange formula'':
\begin{equation}\label{exchange}
  \mathcal{W}^{\alpha'\alpha}(x^{\prime}-j\Lambda(-t)x;e^{\prime},j\Lambda(-t)e)
  = \mathcal{W}^{\alpha\alpha'}%
(x-j\Lambda(t)x^{\prime};e,j\Lambda(t)e^{\prime}).
\end{equation}
Here $j$ is the reflection across the edge of the wedge $W$ which
transforms $W$ into $W^{\prime}$ and $V^+$ into $-V^+$, and
$\Lambda(t)$ is the one-parameter group of Lorentz boosts that leave
$W$ invariant.  Note that $j$ and $\Lambda(t)$ commute.  The matrix
valued function $\Lambda(t)$ is entire analytic in the boost parameter
$t$.  Moreover, for $t$ in the strip $\mathbb{R}+i(0,\pi)$ the
imaginary parts of $j\Lambda(-t)x$, $j\Lambda(-t)e$, $j\Lambda(t)x'$
and $j\Lambda(t)e'$ all lie in $V^+$.  Eq.\ \eqref{exchange} extends
from the boundary at ${\rm Im}\,t=0$ to the whole strip by the analyticity of
the two point function and the Schwarz reflection principle.  The
boundary values for ${\rm Im}\,t=i\pi$ are therefore also identical for both
sides.  Since $j\Lambda(\pm i\pi)$ is the identity matrix, this leads
to the desired stringlike commutativity in the form ${\mathcal
W}^{\alpha'\alpha} (x^{\prime}-x;e^{\prime },e)={\mathcal
W}^{\alpha\alpha'}(x-x^{\prime};e,e^{\prime})$ if $x+\mathbb{R}^{+}e$
and $x^{\prime}+\mathbb{R}^{+}e^{\prime}$ are space-like separated.

The structure of the two-point function also permits the definition of a KMS
(thermal equilibrium) state 
at inverse temperature $\beta$, replacing $M^{\alpha\alpha'}(p;e,e')$ by
\begin{equation}
M^{\alpha\alpha'}_{\beta}(p;e,e^{\prime}) =\left( 1-e^{-\beta
p^{0}}\right)
^{-1}\left[\theta(p^{0})M^{\alpha\alpha'}(p;e,e^{\prime})\right.
\left. 
-\theta(-p^{0})M^{\alpha'\alpha}(-p;e',e)\right]\end{equation} with 
$\theta$ the  step function. The
KMS property is
\begin{equation} M^{\alpha\alpha'}_{\beta}(p;e,e^{\prime}) =e^{\beta p^{0}}
    M^{\alpha'\alpha}_{\beta}(-p;e^{\prime},e).
\end{equation}
The existence of a KMS state is the prerequisite for the
thermalization of a system.  In his discussion of the possible
physical significance of his zero mass infinite spin representations
in \cite{Wig2} Wigner expressed concern about the infinite degeneracy
of each energy level in the one-particle space, that apparently would
lead to a divergence of the partition function in a box.  It is not
clear, however, if such a treatment is legitimate for objects with a
semi-infinite string localization.  This question merits a further
study, including a comparison with the results of \cite{BuJ} on the
thermodynamic properties of conventional quantum fields.

An important open problem in this context is the existence of local
observables in the sense of \cite{Haag}, i.e., operators that are
localized in bounded domains of Minkowski space and relatively local
for the fields.  From the modular duality results of \cite{BGL} it
follows that such operators must be contained in the intersection of
the operator algebras generated by string field operators localized in
wedge domains containing the bounded localization domain, so the
question is whether the intersections of the wedge algebras contain
nontrivial local operators. A sufficent condition based on nuclearity
properties of modular operators has very recently been given 
in \cite{BuL} but it is
restricted to space-time dimensions not larger than two and hence not
applicable in the present case without modifications.

Our results suggest that although string-localized fields are admitted
by the physical principles, they are outside  the realm of Lagrangean 
quantization and hence call for new concepts 
and methods which are more intrinsically rooted in local quantum 
physics \footnote{Looking with the present hindsight 
(of quantum localizability in the
representation theoretical setting) at the early history of quantized 
fields which eventually culminated in renormalized perturbation theory, it
appears as an instance of undeserved luck that the point-like quantum
localizability in Wigner's representation theoretical approach for
particles with finite (half)integer spin/helicity made such a perfect match
with the locality inherent in the classical formalism of local tensor/spinor 
fields.}.  
As a historical remark we point out that already in 1929 
Pascual Jordan made a plea for an instrinsic formulation of QFT 
without {\it
``klassisch-korrespondenzm\"assige Kr\"ucken''} (quasiclassical
crutches) \cite{Jord}. 
The concept of modular localization, that inspired the present 
construction of string localized fields, can be regarded as a modern 
realization of this vision of Jordan.  One of its achievements
is the successful derivation, from first principles, of the recipes of
the bootstrap-formfactor programs for the rich class of $d=1+1$
factorising models \cite{S99, BuL}.  What has been missing up to now
is an example demonstrating beyond doubt that this trans-Lagrangean
point of view is also relevant in  four space-time dimensions. Our string
localized fields provide such examples. 
Furthermore, current work  \cite{M-S-Y} indicates that our 
construction has the potential for further applications. 
Namely, along the same lines string localized fields can be 
constructed also for massive particle types, 
opening the possibility for more general kinds of interactions than
for the usual point-like fields. 
Note in this context that the results of \cite{BGL,BuFr} 
support the viewpoint that localization of quantum fields in 
space--like cones (the idealizations of which are our strings) 
is a natural concept, yet there is so far a lack of rigorous model 
realizations \footnote{apart from non--Lorentz covariant infra--vacua models
as in \cite{Kunhardt} and lattice models as in \cite{Muller,Luscher}.}. 
Now if there is an interacting quantum field with such 
localization, then the corresponding in- and out- fields, in the sense
of LSZ, must be string--localized as well. Hence, our fields 
(in contrast to the usual free fields) may serve as the in- and 
out-fields of such a model. We shall return to this issue 
elsewhere \cite{M-S-Y}. \\

B.S. thanks the ESI, Vienna, and J.Y. the MPI for Physics, Munich and
the Science Institute of the University of Iceland for hospitality
during the completion of this paper.  J.M's work is supported by
FAPESP.\\

\section*{ Appendix: Modular localization}

For convenience of the reader we summarize here some basic definitions
and facts about modular localization, referring to \cite{BGL, Mund1,
F-S} for details.  In a nutshell the idea is that there is a natural
concept of localization of state vectors in space and time that is
defined for certain representations of the Poincar\'e group.  This
concept has its roots in the CPT theorem and an important paper
\cite{BW} of Bisognano and Wichmann.  It is distinct from
Newton-Wigner localization and not associated with any position
operators (that are known to be problematic in relativistic quantum
mechanics).  One first definies localization in space like wedges and
then carries the definition over to more general domains by forming
intersections.

Let $W$ be a  space-like wedge, i.e., a Poincar\'e transform of
the standard wedge $W_{3}\equiv\{x=(x^0,\dots,x^3)\in{\mathbb R}^4:\, |x^0|<x^3\}$. 
To $W$ belongs a one-parameter family $\Lambda_{W}(t)$ of
Lorentz boosts that leave $W$ invariant ($t$ is the rapidity
parameter), and a reflection $j_{W}$ about the edge of the wedge that
maps $W$ into the opposite wedge $W'$.  (The
dependence of these transformations on $W$ was suppressed in Eq.\
(14).)

Let $U$ be a representation of the proper Poincar\'e group ${\mathcal
P}_{+}$ on a Hilbert space ${\mathcal H}$.  It is assumed that $U$ is
unitary on the orthocronous group ${\mathcal P}^{\uparrow}_{+}$ but
antiunitary for the reflections $j_{W}$.  Moreover, the energy
spectrum is assumed to be nonnegative.

For a given wedge $W$, the ``modular operator'' $\Delta_W$ is defined
as the unique positive operator satisfying $
\Delta_W^{it}=U(\Lambda_{W}(-2\pi t))$ for all real $t$.  It is an
unbounded operator (exept in trivial cases) and hence can not be
defined on the whole of ${\mathcal H}$.  The same applies to
$\Delta_W^{1/2}$ which has a natural domain of
definition, $D(W)\subset {\mathcal H}$.  Concretely, $D(W)$ consists
of state vectors $\psi\in{\mathcal H}$ such that $U(\Lambda_{W}({-2\pi
t}))\psi$ can be analytically continued to the strip $0\leq {\rm Im}\,
t\leq \pi$.

Let $J_{W}$ to be the anti--unitary operator representing $j_{W}$. 
The operator $S_W\equiv J_W\,\Delta_W^{1/2}$
(``Tomita conjugation'') 
is defined on $D(W)$ and satisfies $S_{W}^2\subset$id.  
State vectors left invariant under $S_{W}$, 
i.e., belonging to the real subspace
\begin{equation}
{\mathcal K}(W)\equiv \{\phi\in D(W): S_W\phi=\phi \} 
\end{equation}
are said to be {\it localized\/} in the wedge $W$ in the modular
sense.  The space ${\mathcal K}(W)$ is a real Hilbert space with the
real scalar product ${\rm Re}\,(\psi,\phi)$.  Moreover, it satisfies
${\mathcal K}(W)\cap {\rm i}{\mathcal K}(W)=\{0\}$, and ${\mathcal
K}(W)+{\rm i}{\mathcal K}(W)$ is dense in $\mathcal H$.

The localization attribute is justified by the fact that the
symplectic complement \begin{equation}\label{sympl}{\mathcal
K}(W)^\prime\equiv\{\psi:\,{\rm Im}\,(\psi,\phi)=0\,\text{for all}\,
\phi\in {\mathcal K}(W)\}\end{equation} is equal to ${\mathcal
K}(W')$, i.e., the space of state vectors localized in the causal
complement of $W$.  Second quantization allows one to define field
operators $\Phi(\psi)$ on the Fock space over ${\mathcal H}$ such that
$[\Phi(\psi),\Phi(\phi)]={\rm i}\,{\rm Im}\, (\psi,\phi),$ and by Eq. 
\eqref{sympl} $\Phi(\psi)$ and $\Phi(\phi)$ commute if $\psi$ and
${\phi}$ are localized in causally separated wedges.

For more general domains $G\subset{\mathbb R}^4$ one defines the
corrsponding spaces ${\mathcal K}(G)$ of localized vectors as the
intersections of the spaces ${\mathcal K}(W)$ for all wedges $W$
containing $G$.  But while ${\mathcal K}(W)$ is always large in the
sense that ${\mathcal K}(W)+{\rm i}{\mathcal K}(W)$ is dense in
${\mathcal H}$, this is in general not so for ${\mathcal K}(G)$ which
may consist only of the zero vector.  It is a highly nontrivial result
of \cite{BGL} that ${\mathcal K}(G)+{\rm i}{\mathcal K}(G)$ is still
dense if $G$ is a spacelike cone, i.e., a set of the form $x+\{\lambda
y:\,\lambda>0, y\in B\}$ where $x\in{\mathbb R}^4$ and $B$ is an
(arbitrarily small) open set not containig the origin.

The string localized fields (4) realize these
ideas in a concrete setting.  The discussion following Eq.\ (14)  
confirms implicitly that the fields (4) generate states that are 
localized in the modular sense.


\end{document}